\documentclass[dvips]{article}
\usepackage{graphicx}
\usepackage{amssymb}
\usepackage{graphicx}
\usepackage{dcolumn}
\usepackage{amsmath}
\usepackage{lscape}
\usepackage{longtable}

\begin{document}
\textwidth 16cm
\newcommand{\bd}{\begin{document}}
\newcommand{\ed}{\end{document}}
\newcommand{\bc}{\begin{center}}
\newcommand{\ec}{\end{center}}
\newcommand{\bfr}{\begin{flushright}}
\newcommand{\efr}{\end{flushright}}
\newcommand{\lt}{\left}
\newcommand{\rt}{\right}
\newcommand{\vs}{\vspace}
\newcommand{\hs}{\hspace}
\newcommand{\beq}{\begin{equation}}
\newcommand{\eeq}{\end{equation}}
\newcommand{\lb}{\linebreak}
\newcommand{\pb}{\pagebreak}
\newcommand{\mb}{\makebox}
\newcommand{\fb}{\framebox}
\newcommand{\mc}{\multicolumn}
\newcommand{\ben}{\begin{enumerate}}
\newcommand{\een}{\end{enumerate}}
\newcommand{\bit}{\begin{itemize}}
\newcommand{\eit}{\end{itemize}}
\newcommand{\ol}{\overline}
\newcommand{\un}{\underline}
\newcommand{\lefq}{\lefteqn}
\newcommand{\ba}{\begin{array}}
\newcommand{\ea}{\end{array}}
\newcommand{\beqa}{\begin{eqnarray}}
\newcommand{\eeqa}{\end{eqnarray}}
\newcommand{\beqas}{\begin{eqnarray*}}
\newcommand{\eeqas}{\end{eqnarray*}}
\newcommand{\bfg}{\begin{figure}}
\newcommand{\efg}{\end{figure}}
\newcommand{\bds}{\begin{displaymath}}
\newcommand{\eds}{\end{displaymath}}
\newcommand{\btb}{\begin{tabbing}}
\newcommand{\etb}{\end{tabbing}}
\bc {\huge Metric Operator For The Non-Hermitian Hamiltonian Model
and Pseudo-Supersymmetry} \ec

\vs{1cm}

\bc
{\it \"Ozlem Ye\c{s}ilta\c{s}$^{*}${\footnote {e-mail : yesiltas@gazi.edu.tr} and Nafiye Kaplan$^{*}$  \\
$^{*}$Department of Physics, Faculty of Science, Gazi University,
06500 Ankara, Turkey\\
\vspace{.16cm}

}} \ec \vs{1cm}
\begin{abstract}
\noindent We have obtained the metric operator $\Theta=\exp T$ for
the non-Hermitian Hamiltonian model
$H=\omega(a^{\dag}a+1/2)+\alpha(a^{2}-a^{\dag^{2}})$. We have also
found the intertwining operator which connects the Hamiltonian to
the adjoint of its pseudo-supersymmetric partner Hamiltonian for
the model of hyperbolic Rosen-Morse II potential.
\end{abstract}
\noindent {\bf keyword:} metric operator, pt symmetry, pseudo-super-symmetry \\


\section{Introduction}
For over a decade, quantum mechanics in complex domain, i.e. the
field of $\mathcal{PT}$ symmetric quantum mechanics\cite{bender1}
have attracted much more attention \cite{bender2, bender3,Bender,
bender4, bender5, m, jog, ozlem, kh}. On the other hand, studies
on the observations of $\mathcal{PT}$ symmetry in experimental
physics have been attracted much interest \cite{optics1, optics2,
optics3}. The Hamiltonian is $\mathcal{PT}$ symmetric if $[H,
\mathcal{PT}]=0$ and coordinate, momentum operators acting on  the
Hilbert space are affected as $\mathcal{PT} $: $p\rightarrow p$,
$x\rightarrow -x$, $i\rightarrow -i$ and linear operator
$\mathcal{P}$ commutes with the anti-linear operator
$\mathcal{T}$, $[\mathcal{P}, \mathcal{T}]=0$ where
$\mathcal{P}^{2}=\mathcal{T}^{2}=1$. The non-Hermitian quantum
mechanics can be generalized to $\eta$-pseudo-Hermitian quantum
mechanics within the inner products \cite{mos}. If $\hat{A}$ is
the pseudo-Hermitian operator, it satisfies $\eta A
\eta^{-1}=A^{\dag}$ where $\eta=\eta^{\dag}$. The real spectrum of
a non-Hermitian Hamiltonian is connected to a positive definite
metric operator and quasi-Hermiticity\cite{sc} which is related to
$\mathcal{PT}$ symmetry is analyzed in the context of
pseudo-Hermitian theory. Moreover, generalization of
super-symmetric quantum mechanics within the pseudo-Hermiticity
which is pseudo-supersymmetry  has been studied by several authors
\cite{roy11,roy2, roy1, Yes}. The non-Hermitian oscillator  was first
explored in \cite{swan} and in a series of papers the metric
operator and algebraic properties are discussed  \cite{Ass, mus,
quesne, bag, mid}, as well as it has generalized to the solvable models in
quantum mechanics \cite{sin}. In this paper we have introduced a
special form of the Swanson Hamiltonian in \cite{swan} which was studied in \cite{oy1, oy2} and we have
obtained the metric operator. Moreover, the construction
of metrics and Hermitian counterparts associated with $su(2)$
algebra can be found in \cite{Ass}. Then, we have examined the
Hamiltonian operator in a differential form and given the pseudo-Hermiticity
of the Hamiltonian for a specific potential model called as
complex hyperbolic Rosen Morse II.

\section{Pseudo-Hermitian Model}
The reality of the spectrum is obtained with respect to a positive
inner product $<.,.>_{+}$ on the Hilbert space $\mathcal{H}$ in
which $H$ is acting $\eta:\mathcal{H} \rightarrow \mathcal{H} $.
Thus, the pseudo-Hermiticity of the Hamiltonian is given by
\begin{equation}\label{1}
    H^{\dag}=\eta H \eta^{-1}
\end{equation}
where $\eta$ is positive definite, invertible operator that is
related to $\mathcal{C}$ operator by $\eta=\mathcal{CP}$
\cite{Bender} and it may be given as $\eta=e^{-Q}$, especially in
perturbation theory \cite{bender4, bender5}. There may be some
operators $\mathcal{O}$ will also be observables, having real
eigenvalues, such as
\begin{equation}\label{2}
    \mathcal{O}^{\dag}=\eta \mathcal{O} \eta^{-1}.
\end{equation}
If there is a similarity transformation
\begin{equation}\label{3}
    h=\Theta H \Theta^{-1}
\end{equation}
where $\Theta=\sqrt{\eta}$, $h$ is known as the Hermitian
equivalent of $H$ and this is called as the quasi-Hermiticity of $H$
\cite{sc, quesne}. Let us introduce a non-Hermitian model which is
$\mathcal{PT}$ symmetric
\begin{equation}\label{4}
    H=\omega(a^{\dag}a+\frac{1}{2})+\alpha(a^{2}-a^{\dag^{2}})
\end{equation}
where $a, a^{\dag}$ are standard boson operators
\begin{equation}\label{a}
a=\sqrt{\frac{\omega}{2}}\hat{x}+i\frac{\hat{p}}{\sqrt{2\omega}}.
\end{equation}
The spectrum of $H$  is
$E_{n}=(n+1/2)\sqrt{\omega^{2}+4\alpha^{2}}$ which is the spectrum of $h$ too. Now we use the
symbol $\Theta$ for the metric operator and an ansatz for $\Theta$
which is
\begin{equation}\label{5}
    \Theta=\exp T
\end{equation}
where the operator $T$ can be taken as
\begin{equation}\label{6}
    T=\varepsilon a^{\dag}a+\kappa (a^{2}-a^{\dag^{2}})
\end{equation}
where $\epsilon, \kappa$ are constants. If we use the
transformations with $\Theta$ as given below
\begin{equation}\label{7}
    \Theta a \Theta^{-1}=\left(\cosh \theta-\frac{\varepsilon}{\theta}\sinh
    \theta\right) a+\frac{2\kappa}{\theta}\sinh \theta~ a^{\dag}
\end{equation}
\begin{equation}\label{8}
 \Theta a^{\dag} \Theta^{-1}=\left(\cosh\theta+\frac{\varepsilon}{\theta}\sinh\theta\right) a^{\dag}+\frac{2\kappa}{\theta}\sinh\theta~ a
\end{equation}
where we use $\theta=\sqrt{\varepsilon^{2}+4\kappa^{2}}$. To find
the Hermitian equivalent $h$, (\ref{3}) can be expressed as below
\begin{equation}\label{9}
    h=\Theta H
    \Theta^{-1}=f_{1}(\varepsilon, \kappa)(a^{\dag}a+\frac{1}{2})+f_{2}(\varepsilon,
    \kappa)a^{2}-f_{3}(\varepsilon, \kappa)a^{\dag^{2}}.
\end{equation}
Here, $f_{i}(\varepsilon, \kappa), i=1, 2, 3$ are some functions
depend on the Hamiltonian parameters. Now we can use
\begin{equation}\label{10}
    \Theta \Theta^{-1} \Theta a^{\dag} \Theta^{-1} \Theta a \Theta^{-1}=\Theta
    a^{\dag}a \Theta ^{-1}
\end{equation}
and $\Theta a \Theta^{-1}$, $\Theta a^{\dag} \Theta^{-1}$ in order
to obtain $h$. If we perform some algebra, we obtain
\begin{eqnarray}
  f_{1} (\varepsilon, \kappa)&=& \omega \cosh^{2}\theta-\frac{\omega(\varepsilon^{2}-4\kappa^{2})+8\kappa \varepsilon \alpha}{\theta^{2}}\sinh^{2}
  \theta\\ \label{11}
  f_{2}(\varepsilon, \kappa) &=& \frac{2\kappa \omega}\theta L_{-}\sinh\theta+\alpha L^{2}_{-}-\frac{4\kappa^{2}\alpha}{\theta^{2}}\sinh\theta
  \\ \label{12}
  f_{3}(\varepsilon, \kappa) &=& - \frac{2\kappa \omega}\theta L_{+}\sinh\theta+\alpha
  L^{2}_{+}-\frac{4\kappa^{2}\alpha}{\theta^{2}}\sinh\theta \label{13}
\end{eqnarray}
where
\begin{eqnarray}
  L_{-} &=& \cosh\theta-\frac{\epsilon}{\theta} \\ \label{14}
  L_{+} &=&\cosh\theta+\frac{\epsilon}{\theta}. \label{15}
\end{eqnarray}
Then, because $h$ must be Hermitian, considering (\ref{9}) one can
see that $f_{2}(\epsilon, \kappa)=-f_{3}(\epsilon, \kappa)$, thus
we arrive at a condition which is
\begin{equation}\label{16}
    \frac{\tanh^{2}\theta}{\theta^{2}}=\frac{\alpha}{\alpha(4\kappa^{2}-\varepsilon^{2})+2\kappa \omega \varepsilon}
\end{equation}
and this can be thought as the Hermiticity condition. We can use a
parameter $z=\frac{2\kappa}{\varepsilon}$ as used in \cite{mus},
then, $\varepsilon$ is given as
\begin{equation}\label{17}
    \varepsilon=\pm \frac{1}{\sqrt{1-z^{2}}}\tanh^{-1}\sqrt{\frac{\alpha(1-z^{2})}{\omega
    z-\alpha(1-z^{2})}}.
\end{equation}
Using (\ref{a}), we can obtain
\begin{equation}\label{18}
    x=\Theta^{-1} \hat{x}
    \Theta=\left(\cosh\theta+\frac{2\kappa}{\theta}\sinh\theta\right)\hat{x}-i\frac{\varepsilon}{\omega\theta}\sinh\theta\hat{p}
\end{equation}
\begin{equation}\label{19}
    p=\Theta^{-1} \hat{x}
    \Theta=\left(\cosh\theta-\frac{2\kappa}{\theta}\sinh\theta\right)\hat{p}+i\frac{\varepsilon\omega}{\theta}\sinh\theta\hat{x}.
\end{equation}
Thus, the Hermitian equivalent $h$ will have a form
\begin{equation}\label{20}
    h=\lambda_{1}(z)\hat{p}^{2}+\lambda_{2}(z)\hat{x}^{2}
\end{equation}
where $\lambda_{1}(z)$ and $\lambda_{2}(z)$ can be found by using
(\ref{18}), (\ref{19}) and $\varepsilon$ as
\begin{equation}\label{21}
    \lambda_{1}(z)= \frac{\omega}{2}\left(U(1+ \sigma(z))+V\left(\frac{2z(\omega+\alpha z)}{z\omega-2\alpha(1-z^{2})}+2\sqrt{\alpha}\sigma(z)\right)\right)
\end{equation}
and
\begin{equation}\label{22}
   \lambda_{2}(z)=\frac{1}{2\omega}\left(U(1-\sqrt{\alpha}\sigma(z))+V\left(-\frac{4\alpha z}{\omega z-2\alpha (1-z^{2})}+2\sqrt{\alpha}\frac{\sigma(z)}{z}\right)\right)
\end{equation}
where
\begin{equation}\label{23}
    \sigma(z)=z\frac{\sqrt{\alpha(1-z^{2})-z\omega}}{z\omega-2\alpha(1-z^{2})}
\end{equation}
\begin{equation}\label{24}
    U=\omega-\frac{4\alpha^{2}}{\omega z-2\alpha(1-z^{2})}
\end{equation}
\begin{equation}\label{25}
    V=\frac{z\omega(1-\alpha)}{z\omega-2\alpha(1-z^{2})}+(\frac{\omega}{2}-\frac{\alpha}{z})\sqrt{\alpha}\sigma(z).
\end{equation}
Finally, we have obtained $\Theta(z)$ as
\begin{equation}\label{26}
    \Theta(z)=\left(\frac{1+\sqrt{\frac{\alpha(1-z^{2})}{\omega z-\alpha(1-z^{2})}}}{1-\sqrt{\frac{\alpha(1-z^{2})}{\omega
    z-\alpha(1-z^{2})}}}\right)^{\pm
    \frac{1}{2\sqrt{1-z^{2}}}(a^{\dag}a+\frac{z}{2}(a^{2}-a^{\dag^{2}}))}
\end{equation}
If we take $z=0$, then, Hermitian equivalent becomes
\begin{equation}\label{z0}
    h=\frac{\omega}{2} (\omega-2\alpha)\hat{p}^{2}+\frac{1}{2\omega}(\omega-2\alpha)\hat{x}^{2}
\end{equation}
that agrees with the results in \cite{mus} and \cite{swan}.
\section{Pseudo-Supersymmetry}
Let us give (\ref{4}) in terms of the differential operators using
$a, a^{\dag}$ which can be generally given by
\begin{equation}\label{27}
  a=A(x)\frac{d}{dx}+B(x),~~~~a^{\dag}=-A(x)\frac{d}{dx}+B(x)-A(x)'
\end{equation}
where $A(x)$, $B(x)$ are real functions. Now, in terms of
differential operators, (\ref{4}) becomes
\begin{equation}\label{28}
\begin{split}
  H&=-\omega A(x)^{2}\frac{d^{2}}{dx^{2}}+(4\alpha A(x)B(x)-2\omega A(x)A(x)')\frac{d}{dx}\\-
  &(\omega-2 \alpha) A(x)B(x)'-(\omega- 2 \alpha) A(x)' B(x)+\omega B(x)^{2}-\alpha (A(x)A(x)^{''}+A(x)'^{2})+\frac{\omega}{2}.
  \end{split}
\end{equation}
The Hamiltonian can be mapped into a Hermitian operator form by
using a mapping function $\rho$
\begin{equation}\label{29}
  h=\rho H \rho^{-1}
\end{equation}
where
\begin{equation}\label{30}
  \rho=e^{-\frac{2\alpha}{\omega}\int dx \frac{B(x)}{A(x)}}.
\end{equation}
Here we note that $H\Psi=\varepsilon \Psi$, $h \psi=\varepsilon \psi$,
$\Psi=\rho^{-1}\psi$. So we can introduce operator $h$ which is the
Hermitian equivalent of $H$ as
\begin{equation}\label{31}
 h= -\omega\frac{d}{dx}A(x)^{2}\frac{d}{dx}+U_{eff}(x)
\end{equation}
here $U_{eff}(x)$ takes the form
\begin{equation}\label{32}
  U_{eff}(x)=\frac{\omega}{2}-\omega(A(x)B(x))^{'}-\alpha \left((A^{'}(x))^{2}+A(x) A^{''}(x)\right)+(\omega+\frac{4\alpha^{2}}{\omega})B^{2}(x)
  \end{equation}
where the primes denote the derivatives. Then (\ref{31}) can be
mapped into a Schr\"{o}dinger-like form by using
 \begin{equation}\label{32}
   \psi(x)=\frac{1}{A(x)} \Phi(x)
 \end{equation}
Hence, Schr\"{o}dinger-like equation becomes \cite{Yes}
\begin{equation}\label{33}
  -\Phi^{''}(x)+\left(\frac{\omega/2-\varepsilon}{\omega A^{2}(x)}-\frac{(A(x)B(x))^{'}}{A^{2}(x)}+
  \frac{\omega^{2}+4\alpha^{2}}{\omega^{2}}\frac{B^{2}(x)}{A^{2}(x)}
  +\frac{\omega-\alpha}{\omega}\frac{A^{''}(x)}{A(x)}-\frac{\alpha}{\omega}\frac{(A^{'}(x))^{2}}{A^{2}(x)}\right) \Phi=0.
\end{equation}
If we use $A(x)=\cosh x$ and $B(x)=\delta \cosh x$ in above
equation, we obtain the $V(x)$ which is seen as a part in
(\ref{33}) or $-\Phi^{''}(x)+V(x)\Phi(x)=\lambda \Phi(x)$ as
\begin{equation}\label{34}
V(x)=\delta^{2}\frac{\omega^{2}+4\alpha^{2}}{\omega^{2}}-\left(\epsilon-\frac{1}{2}-\frac{\alpha}{\omega}\right)\sec
h^{2}x+2\delta \tanh x
\end{equation}
where $\lambda=\frac{2\alpha}{\omega}$ is taken. This model is
known as hyperbolic Rosen- Morse II potential \cite{suk}. Let us
give a factorization procedure for the model in (\ref{34}), so we
write
\begin{equation}\label{35}
    H_{p}=-\frac{d^{2}}{dx^{2}}+V(x)=-\frac{d^{2}}{dx^{2}}+W(x)^{2}-W(x)'
\end{equation}
\begin{equation}\label{36}
    H=-\frac{d^{2}}{dx^{2}}+\bar{V}(x)=-\frac{d^{2}}{dx^{2}}+W(x)^{2}+W(x)'
\end{equation}
where $W(x)$ is the super-potential, $H_{p}$ is the partner
Hamiltonian of $H$ which is \cite{mos, roy11}
\begin{equation}\label{37}
    \eta H= H^{\dag}_{p} \eta
\end{equation}
where $\eta$ is the intertwining operator and is a linear invertible
operator in pseudo-Hermitian quantum theory. Let us give the super-potential that
has a form as
\begin{equation}\label{38}
    W(x)= a \tanh x+b
\end{equation}
and partner potentials are
\begin{equation}\label{39}
    V(x)=b^{2}+a^{2}-a(a+1) \sec h^{2}x+2ab\tanh x
\end{equation}
\begin{equation}\label{40}
    \bar{V}(x)=b^{2}+a^{2}+a(1-a) \sec h^{2}x+2ab\tanh x
\end{equation}
where
\begin{eqnarray} \label{400}
  a &=& -\frac{4\alpha^{2}\delta^{2}+2\alpha \omega+\omega^{2}(1+\delta^{2}-2\epsilon)}{2\omega^{2}}\pm \delta\frac{\sqrt{
  (\delta^{2}-4)\omega^{4}+8\alpha^{2}\delta^{2}\omega^{2}+16\alpha^{4}\delta^{2}}}{2\omega^{2}}\\ \label{401}
  b &=&\frac{\delta}{a}
\end{eqnarray}

Let us choose the parameter $b$ as $b\rightarrow i b$, then we
have complex partner potentials
\begin{eqnarray}
  V(x) &=& -b^{2}+a^{2}-a(a+1) \sec h^{2}x+2iab\tanh x \\
  \bar{V}(x) &=&- b^{2}+a^{2}+a(1-a) \sec h^{2}x+2iab\tanh x.
\end{eqnarray}
And the adjoint of $H_{p}$ is
\begin{equation}\label{41}
    H^{\dag}_{p}=-\frac{d^{2}}{dx^{2}}+V^{\dag}(x)=-\frac{d^{2}}{dx^{2}}-b^{2}+a^{2}+a(a+1) \sec h^{2}x-2iab\tanh x
\end{equation}
where $ W(x)$ can be given as $W(x)=a \tanh x+ib$. Now, let us
find the $\eta_{1}$ which intertwines $H$ and $H_{p}$ as
$\eta_{1}H=H_{p}\eta_{1}$. Then, we find $\eta_{1}$  as given
below
\begin{equation}\label{42}
  \eta_{1}(x) =\frac{d}{dx} -a\tanh x+ib
\end{equation}
On the other hand, we can give
\begin{equation}\label{43}
    \eta_{2}H_{p}=H^{\dag}_{p}\eta_{2}
\end{equation}
which means that $H_{p}$ is $\eta_{2}$-pseudo-Hermitian
\cite{roy11}. Then, we may give $\eta_{2}$ as
\begin{equation}\label{44}
    \eta_{2}=\frac{d}{dx}-i\frac{a+1}{2b} \sec h^{2} x
\end{equation}
and finally one can give the operator $\eta$ as
\begin{equation}\label{et}
    \eta=\left(\frac{d}{dx}-i\frac{a+1}{2b} \sec h^{2} x \right)\left( \frac{d}{dx}-a\tanh
    x+ib\right).
\end{equation}
The energy spectrum and the exact solutions of the (\ref{35}) can
be given as \cite{levai}
\begin{eqnarray}
  \lambda_{n} &=& -(a-n)^{2}+\frac{a^{2}b^{2}}{(a-n)^{2}} \\
  \Phi_{n} &=& N (1-\tanh x)^{\frac{c_{1}}{2}}(1+\tanh x)^{\frac{c_{2}}{2}}P^{(c_{1}, c_{2})}_{n}(\tanh x)
\end{eqnarray}
where $Re~ (c_{1}) >0, Re~ (c_{2}) >0$
\begin{equation}\label{45}
    n=0,1,...n_{max}, ~~~~\frac{c_{1}+c_{2}}{2}-a=-n,~~~~~c_{1n}=a-n+\frac{i\lambda}{a-n},~~~~c_{2n}=a-n-\frac{i\lambda}{a-n}.
\end{equation}
$H$ and $H^{\dag}_{p}$ are pseudo-super-partner Hamiltonians with the energy spectrum given above. In \cite{levai}, the pseudo-norm for the wave-function was given
in detail. One can look at \cite{levai} for obtaining the
normalization constant $N$. Then, we gave probability density
graphs using the Hamiltonian parameters in figures $1, 2, 3$.

\section{Conclusion}
In conclusion, the metric operator is constructed for the model in
(\ref{4}). Taking $z=0$ gives the parallel results with
\cite{mus}. In \cite{bag}, the authors studied a generalized quantum condition for the Swanson Hamiltonian and
the symmetric nature of the Hamiltonian with respect to the parameters $\alpha$ and $\beta$. The model studied in this paper is
the special case of the general frame related to the Hamiltonian in \cite{bag} when the parameters are taken as $\alpha=-\beta$.
In our study we have used an operator (\ref{5}) which is the metric operator where we take the operator $T$ similar to the
non-Hermitian Hamiltonian. The generalized  Bogoliubov transformations using a general transfomation operator and diagonalization gave the Hamiltonian in Harmonic oscillator form \cite{swan}, and in this paper  after obtaining the metric operator we have discussed a special case (\ref{z0})
which is related to the harmonic oscillator.  We have also studied the exact solvability of the
model as giving the bosonic operators in terms of differential
operators and according to the special choices of the functions in
the first order differential operators we have given a special
potential model. Using the concepts of pseudo-super-symmetry, we
have obtained the pseudo-super-symmetric partners of the hyperbolic
Rosen-Morse II potential in case of taking the one of the
potential parameter $b$ as $i b$. The operator $\eta$
which leads to the real spectrum is obtained. The spectrum and
wave-functions are given in terms of Hamiltonian parameters. We
have given the probability density graphs for $n=1$, $n=2$ and
$n=8$. We have seen that the bound-states of the model depend on
the parameters of the Hamiltonian which can also be seen from the
graphs. In these graphs, the normalization constant is used as
given in \cite{levai}.


\begin{figure}[!htb]
\centering
\includegraphics[scale=.7]{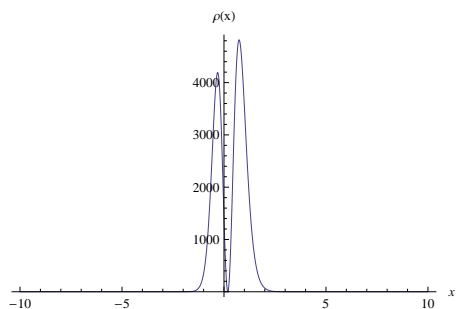}
\caption{Graph of $\rho=|\Phi|^{2}$ for the parameters: $\alpha =
2; n = 1; \omega = 3; \delta = 10; \epsilon = 5$.}
\label{fig:digraph}
\end{figure}

\begin{figure}[!htb]
\centering
\includegraphics[scale=.7]{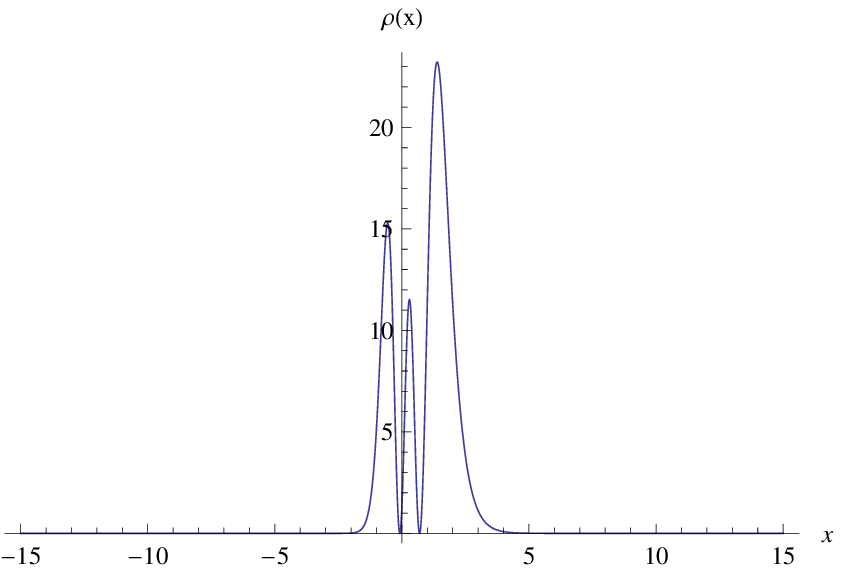}
\caption{Graph of $\rho=|\Phi|^{2}$ for the parameters: $\alpha = 2; n = 2; \omega =
3; \delta = 10; \epsilon = 4$. } \label{fig:digraph}
\end{figure}

\begin{figure}[!htb]
\centering
\includegraphics[scale=.7]{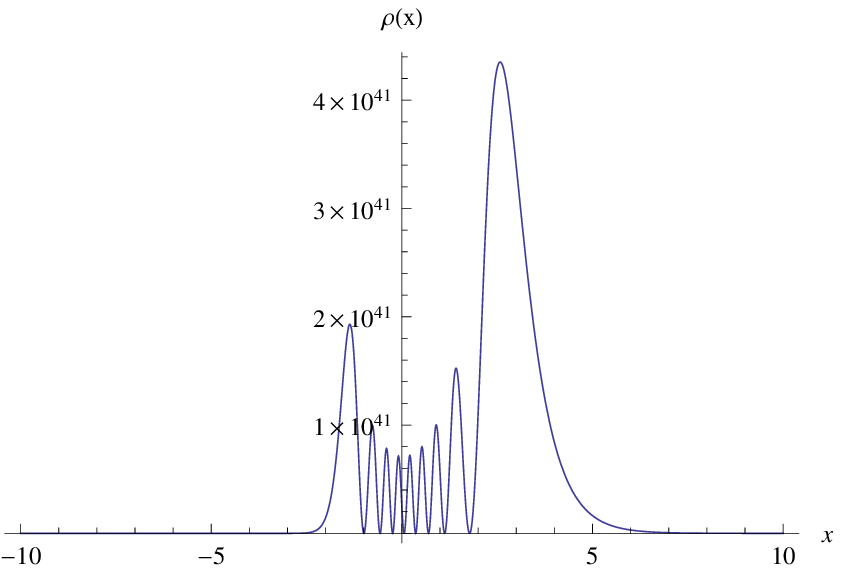}
\caption{Graph of $\rho=|\Phi|^{2}$ for the parameters: $\alpha = 2; n = 8; \omega =
 3; \delta = 10; \epsilon = 4$. } \label{fig:digraph}
\end{figure}

\end{document}